\documentstyle[12pt,epsfig]{cernart} 
\makeatletter
\def\thebibliography#1{\section*{\refname\@mkboth
 {\uppercase{\refname}}{\uppercase{\refname}}}\list
 {\@biblabel{\arabic{enumi}}}{\settowidth\labelwidth{\@biblabel{#1}}%
 \leftmargin\labelwidth
 \advance\leftmargin\labelsep
 \usecounter{enumi}
 \def\theenumi{\arabic{enumi}}}%
 \def\newblock{\hskip .11em plus.33em minus.07em}%
 \sloppy\clubpenalty4000\widowpenalty4000
 \sfcode`\.=1000\relax}
\makeatother

\begin {document}
\begin {titlepage}
\docnum {CERN--PPE/94--57}
\date {April $8^{th}$, 1994} 
\vspace{1cm}
\title {MEASUREMENT OF THE SPIN-DEPENDENT STRUCTURE FUNCTION
\protect$g_1(x)$ OF THE PROTON}

\vspace{2cm}
\collaboration {The Spin Muon Collaboration (SMC)}
\vspace {4cm}
\begin{abstract}
\vspace {1cm}
We have measured the spin-dependent structure function $g_1^p$ of the
proton in deep inelastic scattering of polarized muons off polarized
protons, in the kinematic range $0.003<x<0.7$ and 
$1\,\mbox{GeV}^2<Q^2<60\,\mbox{GeV}^2$. Its first moment, $\int_0^1 g_1^p(x) dx $,
is found to be $0.136 \pm 0.011\,(\mbox{stat.})\pm 0.011\,(\mbox{syst.})$
at $Q^2=10\,\mbox{GeV}^2$. This value is smaller
than the prediction of the Ellis--Jaffe sum rule by two standard
deviations, and is consistent with previous measurements.
A combined analysis of all available proton, deuteron and neutron
data confirms the Bjorken sum rule to within $10\%$
of the theoretical value.
\end{abstract}
\vspace {2cm}
\submitted {Submitted to Physics Letters B}
\vspace {2cm}

\newpage

\input{authors}
\end {titlepage}
\newpage

The spin dependent structure functions of the nucleon, $g_1$ and
$g_2$, can be measured in polarized deep inelastic lepton-nucleon
scattering \cite{HK83}. Measurements of $g_1$ for the proton and
the neutron allow us to test a fundamental QCD sum rule, derived by
Bjorken \cite{Bj66}, and to study the internal spin structure of
the nucleon. Ellis and Jaffe~\cite{EJ74} have derived sum rules for
the proton and for the neutron, under the assumptions that the
strange sea is unpolarized and that SU(3) symmetry is valid for the
baryon octet decays.

First measurements of $g_1^p$ were performed by experiments at SLAC
(E80 and E130~\cite{Al76}) and at CERN (EMC \cite{As88}). The analysis
of these data \cite{VWH,As88} showed a deviation from the Ellis--Jaffe
prediction, with the implications that the total contribution of quark
spins to the nucleon spin is small and that the strange sea is
negatively polarized. Recently, two experiments have measured $g_1^d$
from polarized muon-deuteron (SMC \cite{SMC93} at CERN) and $g_1^n$
from polarized electron-$^3$He scattering (E142 \cite{E142} at SLAC).
The conclusions from these two experiments appeared to be at variance.
However, combined analyses~\cite{EK93,CR93,ANR93,SMC93B} showed that
the experimental data agree in the kinematic region of overlap, and
emphasized that the conclusions are very sensitive to the small-$x$
extrapolation of $g_1(x)$ and to higher order and higher twist QCD
corrections. Additional data are required to provide a more stringent
test of the sum rules and to clarify the contribution of the quark
spins to the nucleon spin.

In this paper, we report the results of a new measurement of $g_1^p$ at
CERN, where longitudinally polarized muons were scattered from
longitudinally polarized protons in the kinematic range
$1$\,GeV$^2<Q^2<60$\,GeV$^2$ and $0.003<x<0.7$. The experiment is
similar to the previous SMC experiment with a deuteron target
\cite{SMC93}.

The positive muon beam had an intensity of $4 \times 10^7$ muons/spill
with a spill time of 2.4~s, a period of 14.4\,s, and an average muon
energy of 190 GeV. The beam polarization was determined from the the
shape of the energy spectrum of positrons from the decay
$\mu^+\rightarrow e^+\nu_{e}\bar{\nu}_{\mu}$. The polarimeter is
described in Ref.~\cite{SMC94A}. The polarization was measured to be
$P_{\mu}= -0.803 \pm 0.029\,(\mbox{stat.}) \pm 0.020 \,(\mbox{syst.})$,
in good agreement with Monte-Carlo simulations of the beam transport
\cite{Do94}.

A new polarized target was built for this experiment. Its design is
similar to that used in the earlier EMC proton~\cite{As88} and SMC
deuteron~\cite{SMC93} experiments. 
The target consists of an upstream and a downstream cell, each 60~cm
long and 5~cm in diameter, separated by 30~cm, and with opposite
longitudinal polarizations.
The target
material was butanol with about 5\% of water, in which paramagnetic
complexes \cite{KR91} were dissolved, resulting in a concentration of
$7.2 \times 10^{19}$ unpaired electrons per cm$^3$. The material was
frozen into beads of about 1.5~mm diameter.

A new superconducting magnet system \cite{DA91} and a new $^3$He/$^4$He
dilution refrigerator were constructed. The magnet system consists of a
solenoid, 16 correction coils, and a dipole. The solenoid provides a
magnetic field of 2.5\,T with its axis aligned along the beam direction
and with an homogeneity of $2 \times 10^{-5}$ throughout the target
volume.  The dipole magnet provides a magnetic field of up to 0.5\,T in
the vertical direction. The dilution refrigerator achieved a
temperature of about 0.3~K with a cooling power of 0.3~W when
polarizing. The typical temperature in frozen spin operation was below
60~mK.

Protons were polarized by dynamic nuclear polarization (DNP). This was
obtained by applying microwave power near the resonance frequency of
the paramagnetic molecules. To achieve opposite proton polarizations in
the two target cells simultaneously, we used slightly different
microwave frequencies. In addition, frequency modulation of the
microwaves reduced the polarization buildup time by about 20\% and
increased the maximum polarization by 6\%. The mean polarization 
throughout the data-taking was $0.86$, with a maximum value of 0.94.

The measurement of the proton polarization, $P_T$, was performed with
10 coils along the target using continuous-wave NMR with series Q-meter
circuits \cite{CO93,NMR}. The NMR signals were calibrated by measuring
the thermal equilibrium signals at different temperatures around 1~K, where the natural
polarization ($\simeq 0.25$\%) is known from the Curie law.  The
thermal equilibrium signals were corrected for systematic effects
including a small change in size with the field polarity and the
contamination with background signals. The signals were also corrected
for Q-meter non-linearity effects present at large polarizations. The
relative accuracy of the polarization measurement was 3\%.

The spin directions were reversed every 5 hours with only small losses
of polarization and running time, by rotating the
magnetic field direction using a superposition of the solenoid and the dipole
fields. In addition, the spin polarization in each target cell was
reversed via DNP once a week. During spin reversals by field rotation,
the field was made slightly inhomogeneous to avoid depolarization due
to superradiance \cite{Ki88}.

The momentum of the incident muon was measured using a bending magnet
upstream of the target. Its track was reconstructed from hits in
scintillator hodoscopes and proportional chambers. The trajectory and
the momentum of the scattered muon were determined from hits in a total of 150
planes of proportional chambers, drift chambers and streamer tubes
located upstream and downstream of the forward spectrometer magnet
(FSM). The large number of planes minimized the effect of individual
plane inefficiencies on the overall track reconstruction efficiency.
The scattered muon was identified by having traversed a 2~m thick
hadron absorber made of iron. Incident and scattered muon tracks
determined the interaction vertex with an average resolution of 30~mm
(0.3~mm) in the direction parallel (perpendicular) to the beam
direction.

The readout of the apparatus was triggered by coincident hits in three
large scintillator hodoscopes, one located just downstream of the FSM
and two located downstream of the hadron absorber. A dedicated trigger
for events with small scattering angles used hodoscopes with finer
segmentation close to the beam and covered mainly the small $x$ range.

Cuts were applied to minimize smearing effects, to limit the size of
radiative corrections, to reject muons originating from the decay of
pions produced in the target, to ensure that the beam flux was
the same for both target cells and to ensure proper separation of
events originating from the upstream and downstream target cells.
After cuts, the data sample amounted to $3.1\times10^6$ events and
$1.3\times10^6$ events for the large and small angle triggers,
respectively.

The  virtual-photon proton asymmetry $A_1^p$ is related to
the measured muon-proton asymmetry $A^p = (\sigma^{\uparrow\downarrow} -
\sigma^{\uparrow\uparrow}) / ( \sigma^{\uparrow\downarrow} +
\sigma^{\uparrow\uparrow} ) $ by \cite{HK83}
\begin{eqnarray}
A_1^p = \frac{\sigma_{1/2}- \sigma_{3/2}}{\sigma_{1/2}+\sigma_{3/2}} =
\frac{A^p}{D} - \eta A_2^p,
\end{eqnarray}
where 1/2 and 3/2 are the total spin projections in the direction of
the virtual photon. The depolarization factor $D$ and the coefficient
$\eta$ depend on the event kinematics. In addition, $D$ depends on the
unpolarized structure function $R(x,Q^2)$, which was taken from a
global fit of the SLAC data~\cite{SLAC90}. The  asymmetry $A_2^p$
arises from the interference between transverse and longitudinal
virtual photon polarizations and is constrained by the positivity
limit $|A_2^p| \leq \sqrt{R}$.  We have measured $A_2^p$ in a dedicated
experiment, where the dipole field was used to hold the proton
polarization in a direction perpendicular to the beam. We found $A_2^p$
to be compatible with zero within a statistical uncertainty of 0.20,
which is a stronger constraint than the one imposed by the positivity
limit. In addition, since the coefficient $\eta$ is small in the
kinematic range covered by our experiment, we neglected the term $\eta
A_2^p$ and included its possible effect in the systematic error.

The asymmetry $A_1^p$ is extracted from combinations of data sets taken
before and after a polarization reversal. Since we take data simultaneously
with oppositely polarized cells, the incident muon flux, the amount of
material in the target cells and the absolute value of the spectrometer
acceptances, $a_u$ and $a_d$, cancel in the determination of $A_1^p$.
The subscripts $u$ and $d$ refer to the upstream and downstream
target cells, respectively.  The only assumption in deriving $A_1^p$ is
that the ratio $r=a_u/a_d$ remains constant within the typical period
of time between two polarization reversals ($\Delta t\simeq 5$~hours).
A time dependence of $r$ leads to a false asymmetry of
\begin{equation} 
\Delta A_1^p = \frac{1}{4 f P_\mu P_T D} \frac{\Delta r}{r},
\end{equation}
where the dilution factor $f$ is the fraction of the event yield from
protons of hydrogen in the target  ($f \simeq 0.12$). In order
to estimate the uncertainty due to this effect, we have studied the time
dependences of all detector efficiencies; we then reprocessed the data
after artificially imposing on the whole sample the largest of the
variations observed within two polarization reversals.  We also
reanalyzed the data ignoring the information from a fraction of the
planes in our chambers. In this way, we artificially reduced the
redundancy of the spectrometer and became more sensitive to time
dependences. Finally, we divided the data into different subsets
according to a variety of criteria (e.g. data-taking periods, radial
vertex position, events reconstructed in different parts of the
spectrometer) and looked for disagreements between the different
samples. From these studies we concluded that
$\Delta r/r < 5 \times 10^{-4}$, corresponding to a false asymmetry
$\Delta A_1^p < 5\times 10^{-3}$.

Spin-dependent radiative corrections to $A_1^p$ were calculated using
the approach of Ref.~\cite{Ak93}.  They were found to be small over the
whole kinematic range. The uncertainty in the radiative corrections
arises predominantly from uncertainties in the structure functions used as
input. Asymmetries arising from electroweak interference are
negligible in the $Q^2$ range of this experiment.

\begin{figure}[H]
 \begin{center}
 \mbox{ \epsfig{file=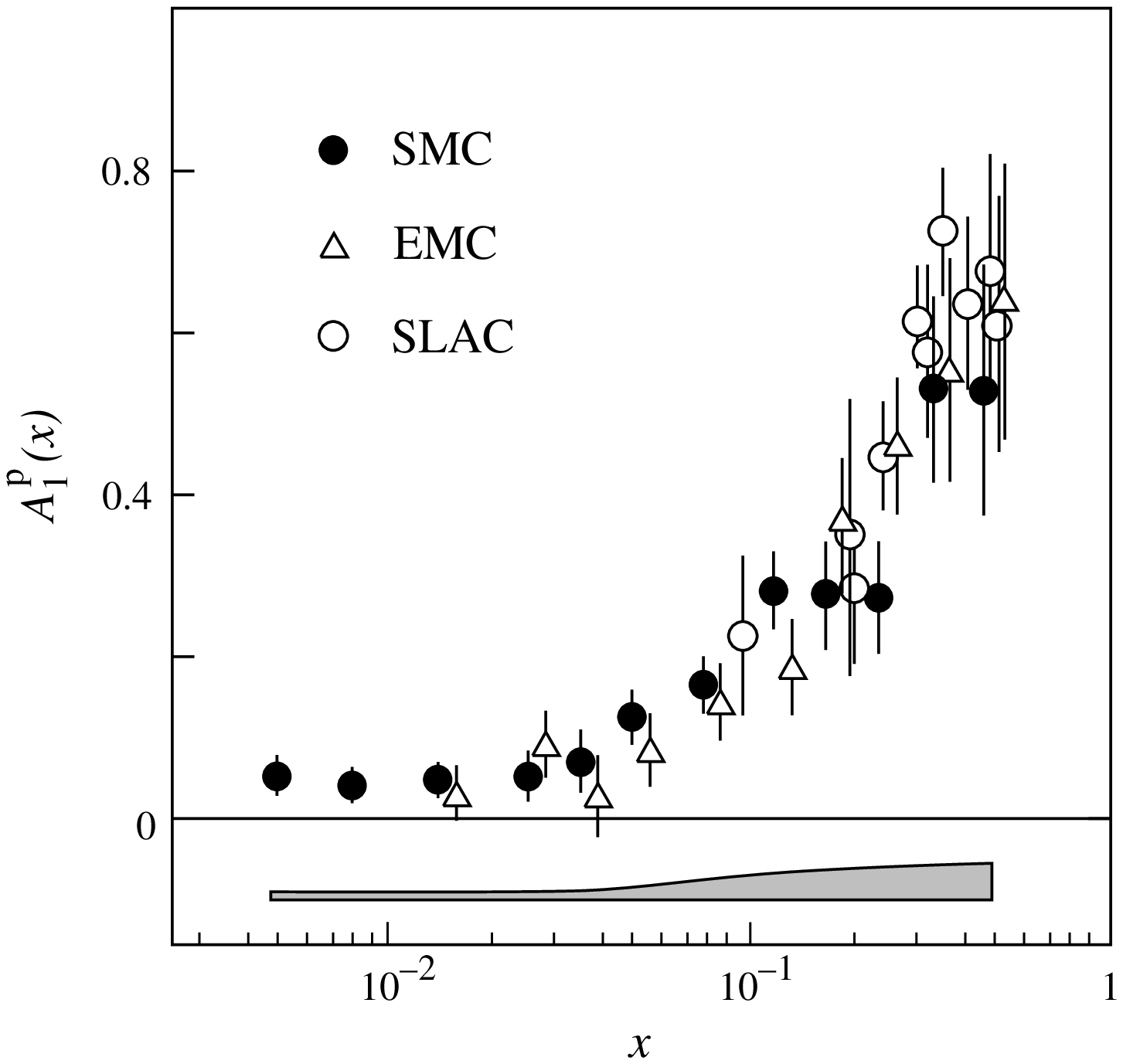,width=10cm}}
 \end{center}
 \caption {\em
 The virtual-photon proton cross section asymmetry $A_1^p$ as a
 function of the Bjorken scaling variable $x$. Only statistical errors
 are shown with the data points. The size of the systematic errors for
 the SMC points is indicated by the shaded area.}
 \label{fig1}
 \vspace{5mm}
 \end{figure}

The results for $A_1^p$ for each $x$ bin at the respective mean $Q^2$
are given in Table~\ref{table1}, and are shown in Fig.~\ref{fig1}.  Sources
of systematic errors include the uncertainties in the beam and target
polarizations, the structure function $R$, the dilution factor $f$, the
radiative corrections, the momentum measurement, the kinematic smearing
corrections, the stability in time of the acceptance ratio, and the
neglect of $A_2$.  The different systematic errors were combined in
quadrature.

The spin structure function $g_1^p$ was evaluated from the average
asymmetry $A_1^p$ in each~$x$ bin using the relation
\begin{equation}
 g_1^p(x,Q^2) = \frac{A_1^p(x,Q^2) F_2^p(x,Q^2)}{2x[1+R(x,Q^2)]}.
\label{Eq_g1}
\end{equation}
The unpolarized structure function, $F_2^p(x,Q^2)$, was taken from the
NMC parametrization \cite{NMC92C}. The uncertainty is typically
3\% to 5\%. The lowest $x$ bin is outside the kinematic region covered
by the NMC data, but we have verified that their
parametrization extrapolates smoothly to the HERA
data~\cite{HERA}, and estimated the corresponding uncertainty to be 15\%.
The structure function $g_1^p$ is practically independent of $R$ due
to cancellations between the implicit $R$ dependences in $D$ and $F_2$
and the explicit one in Eq.~\ref{Eq_g1}. Results for $g_1^p$ are
given in Table~\ref{table1} and Fig. 2.

To evaluate the integral $\int{g_1^p(x,Q^2_0)dx}$ at a fixed $Q^2$, 
we recalculated $g_1^p$ at $Q^2_0=10\,\mbox{GeV}^2$, which represents
an average value for our data. Using Eq.~\ref{Eq_g1}, we obtained
$g_1^p(x,Q^2_0)$ in each bin assuming $A_1(x,Q^2)$ to be independent of
$Q^2$.
This assumption is consistent with our data, with previous experimental
results for both the proton~\cite{As88} and deuteron~\cite{SMC93}, and
with recent theoretical calculations~\cite{ANR93}.
The values of $g_1^p(x,Q^2_0)$ are shown in Table~\ref{table1}.

The integral over the measured $x$ range is
\begin{equation}
 \int_{0.003}^{0.7}g_1^p (x,Q_0^2)dx = 0.131 \pm 0.011 \pm 0.011.
\end{equation}
Here, and in the following, the first error is statistical and the
second is systematic. The contributions to the systematic error are
detailed in Table~\ref{syserr}.
 
To estimate the integral for $x > 0.7$, we take $A_1^p = 0.7 \pm 0.3$
for $0.7 < x < 1.0$, which is consistent with the bound $A_1 < 1$,
and also with the result from perturbative QCD~\cite{Br94} that 
$A_1 \rightarrow 1$ as $x \rightarrow 1$.
This contribution amounts to $0.0015 \pm 0.0007$. The integral
$\int_{x_m}^1g_1^p(x)dx$ as a function of the lower integration
limit, $x_m$, is shown in Fig.~3. The contribution to the integral
from the unmeasured region $x < 0.003$ was evaluated 
assuming a Regge-type dependence $g_1^p(x) =${\it constant}
\cite{He73}, that we fit to our two lowest $x$ data points.  We obtain
$\int_0^{0.003}g_1^p(x)dx = 0.004 \pm 0.002 (\mbox{stat.})$. We
increase this error to 0.004 so that it covers the results obtained
when either the lowest $x$ point or the three lowest $x$ points are used to
determine the extrapolation. This range also covers the results obtained
using the general form of Regge dependence $g_1^p(x) \propto
x^\alpha$, with $0<\alpha<0.5$~\cite{He73}.  Although $g_1$ shows
a tendency to increase at low $x$, Table 1, we do not consider the
trend significant enough to call into question the validity of Regge
behavior.

The result for the first moment of $g_1^p(x)$ at $Q^2_0 = 10$\,GeV$^2$ is
\begin{equation}
 \Gamma_1^p(Q^2_0) =
 \int_{0}^{1}g_1^p (x,Q_0^2)dx = 0.136 \pm 0.011 \pm 0.011.
\end{equation}
The Ellis--Jaffe sum rule, including first order QCD corrections~\cite{Ko79},
predicts $\Gamma_1^p = 0.176 \pm 0.006$ for $\alpha_s(10$\,GeV$^2) =
0.23 \pm 0.02$,
corresponding to $\alpha_s(M_Z^2) = 0.113 \pm 0.004$ \cite{PD92} and
four quark flavors.
Our measurement is two standard deviations below this value.

The first moment $\Gamma_1^p$ can be expressed in terms of the proton
matrix element of the flavor singlet axial vector current
$a_0$~\cite{As88} and the SU(3) coupling constants $F$ and
$D$~\cite{Bo83}.  We obtain $a_0 = 0.18 \pm 0.09 \pm 0.09$.  In the
quark-parton model, $a_0$ is proportional to $\Delta \Sigma=\Delta u
+\Delta d +\Delta s$, the sum of the quark spin contributions to the
nucleon spin.  Our result corresponds to
\begin{equation}
 \Delta \Sigma=0.22 \pm 0.11 \pm 0.11
\end{equation}
and
\begin{equation}
 \Delta s = -0.12 \pm 0.04 \pm 0.04.
\end{equation}
We thus find that only a small fraction of the nucleon spin is due to
the helicity of the quarks, and that the strange sea is negatively
polarized. 

Our results are in good agreement with the previous measurements of
E80/E130 and the EMC. A test of consistency of the experimental
asymmetries $A_1^p(x)$ from all experiments yields  $\chi^2 =
14.6$ for 15 degrees of freedom. To compare $g_1^p$, values we apply to the EMC
asymmetries the same $F_2^p$ parametrization that we use in the
present work. An evaluation of the integral over the $x$ range
common to both experiments, at $Q^2_0 = 10$\,GeV$^2$, yields
$\int_{0.01}^{0.7}g_1^p(x) dx = 0.124 \pm 0.013 \pm 0.019$ for the EMC and
$\int_{0.01}^{0.7}g_1^p(x) dx = 0.118 \pm 0.010 \pm 0.010$ for our data.
In the range $0 < x < 0.01$, the extrapolation of the EMC
data gives $0.003 \pm 0.003$,
while our two lowest $x$ points and our extrapolation 
give $\int_{0}^{0.01}g_1^p(x) dx =0.018 \pm 0.006$.

For a common evaluation of $\Gamma_1^p$ from all existing data, we
combine our results on $A_1^p(x)$ with those of E80/E130 and EMC. The
extrapolations are recalculated from the combined asymmetries following
the methods described above. The treatment of the systematic errors
takes into account that some of them are correlated between the
different experiments. This yields
\begin{equation}
\Gamma_1^p(10\,\mbox{GeV}^2) = 0.145 \pm 0.008 \pm 0.011\ \ \ \ 
\mbox{\it (All\ proton\ data)},	
\label{Eq_allp}
\end{equation}
which is two standard deviations below the Ellis--Jaffe prediction.
From this result, we obtain
$\Delta \Sigma = 0.30 \pm 0.07 \pm 0.10$ and 
$\Delta s = -0.09 \pm 0.02 \pm 0.04$.


We now turn to a test of the Bjorken sum rule~\cite{Bj66}, using all available proton, neutron and deuteron data. We do this
test at $Q^2=5$\,GeV$^2$ in order to avoid a large $Q^2$ evolution of
the SLAC-E142 neutron data, which have an average $Q^2=2$\,GeV$^2$.  A fit to
$\Gamma_1^p$ (Eq.~\ref{Eq_allp}), $\Gamma_1^n$~\cite{E142} and $\Gamma_1^d$~\cite{SMC93},
reevaluated at $5$\,GeV$^2$ under the assumption that the asymmetries
$A_1$ are independent of $Q^2$, yields
\begin{equation} \label{exp-bjorken}
\Gamma_1^p-\Gamma_1^n = 0.166 \pm 0.017  \ \ \ \ \ \ (Q^2=5\,\mbox{GeV}^2),
\end{equation}
where statistical and systematic errors are combined in quadrature.
When one uses the available deuteron and proton data to replace the
extrapolation on the neutron data, as discussed in Ref.~\cite{SMC93B},
one obtains $\Gamma_1^n = -0.069\pm 0.025$ and
\begin{equation} \label{exp-bjorken2}
\Gamma_1^p-\Gamma_1^n = 0.203 \pm 0.029  \ \ \ \ \ \ (Q^2=5\,\mbox{GeV}^2),
\end{equation}
with a larger error due to the limited statistics in the deuteron experiment.
The theoretical prediction, including perturbative QCD corrections
up to third order in $\alpha_s$ \cite{LV91}, gives
\begin{equation} \label{th-bjorken}
\Gamma_1^p-\Gamma_1^n = 0.185 \pm 0.004 \ \ \ \ \ \ \ \ \mbox{\it (Theory)} \ \ \ \ \ \ (Q^2=5\,\mbox{GeV}^2),
\end{equation}
which is in agreement with the above experimental results.  Higher-twist
effects may contribute, especially at low $Q^2$ \cite{BBK90,JU93}, and
have been estimated \cite{ANR93,JU93} to change
$\Gamma^p_1 - \Gamma^n_1$ by about 2\%, but even the sign is uncertain.
We have not taken these contributions into account. 

In summary, we have presented a new measurement of the proton
spin dependent structure function $g_1^p$. The measured asymmetries are in
agreement with those of the earlier E80/E130 and EMC experiments,
but systematic errors have been significantly reduced and the kinematic
region has been extended down to $x=0.003$.
 
The first moment of the spin dependent structure function $g_1^p$,
evaluated from our own data, is two standard deviations below the prediction
of the Ellis--Jaffe sum rule. In the quark parton model, this result
implies that the contribution of the quark spins to the proton spin
is $0.22\pm0.15$. The Bjorken sum rule is now confirmed, at the one standard 
deviation level, to within $10\%$ of its theoretical value.

 \begin{table}[p] \centering
\begin{tabular}{|c|c|c|c|c||c|}
\hline \hline
 $x$-range &$\langle x \rangle$&$\langle Q^2 \rangle$&$A_1^p$& $g_1^p$& $g_1^p$ \\
& & $({\rm GeV}^2)$ & & &$(Q^2$$=$$10\,{\rm GeV}^2)$ \\
\hline \hline
 $ 0.003$--$0.006 $ & $0.005$ & $ 1.3 $ & $ 0.053$$\pm$$0.025$$\pm$$0.007 $ & $ 1.34$$\pm$$0.62$$\pm$$0.27 $ & $ 2.51$$\pm$$1.16$$\pm$$0.69$\\        
 $ 0.006$--$0.010 $ & $0.008$ & $ 2.1 $ & $ 0.042$$\pm$$0.024$$\pm$$0.005 $ & $ 0.73$$\pm$$0.42$$\pm$$0.11 $ & $ 1.15$$\pm$$0.65$$\pm$$0.19$\\        
 $ 0.010$--$0.020 $ & $0.014$ & $ 3.7 $ & $ 0.048$$\pm$$0.022$$\pm$$0.005 $ & $ 0.52$$\pm$$0.24$$\pm$$0.06 $ & $ 0.67$$\pm$$0.31$$\pm$$0.09$\\        
 $ 0.020$--$0.030 $ & $0.025$ & $ 6.0 $ & $ 0.050$$\pm$$0.031$$\pm$$0.005 $ & $ 0.34$$\pm$$0.21$$\pm$$0.03 $ & $ 0.39$$\pm$$0.24$$\pm$$0.04$\\        
 $ 0.030$--$0.040 $ & $0.035$ & $ 8.1 $ & $ 0.069$$\pm$$0.039$$\pm$$0.006 $ & $ 0.35$$\pm$$0.20$$\pm$$0.03 $ & $ 0.37$$\pm$$0.21$$\pm$$0.03$\\        
 $ 0.040$--$0.060 $ & $0.049$ & $10.8 $ & $ 0.124$$\pm$$0.034$$\pm$$0.009 $ & $ 0.46$$\pm$$0.13$$\pm$$0.03 $ & $ 0.46$$\pm$$0.13$$\pm$$0.03$\\        
 $ 0.060$--$0.100 $ & $0.077$ & $15.5 $ & $ 0.161$$\pm$$0.035$$\pm$$0.012 $ & $ 0.38$$\pm$$0.08$$\pm$$0.03 $ & $ 0.36$$\pm$$0.08$$\pm$$0.03$\\        
 $ 0.100$--$0.150 $ & $0.122$ & $22.1 $ & $ 0.275$$\pm$$0.047$$\pm$$0.019 $ & $ 0.40$$\pm$$0.07$$\pm$$0.03 $ & $ 0.38$$\pm$$0.07$$\pm$$0.03$\\        
 $ 0.150$--$0.200 $ & $0.172$ & $28.5 $ & $ 0.273$$\pm$$0.067$$\pm$$0.020 $ & $ 0.26$$\pm$$0.06$$\pm$$0.02 $ & $ 0.25$$\pm$$0.06$$\pm$$0.02$\\        
 $ 0.200$--$0.300 $ & $0.241$ & $36.3 $ & $ 0.267$$\pm$$0.070$$\pm$$0.022 $ & $ 0.16$$\pm$$0.04$$\pm$$0.01 $ & $ 0.16$$\pm$$0.04$$\pm$$0.01$\\        
 $ 0.300$--$0.400 $ & $0.342$ & $46.4 $ & $ 0.529$$\pm$$0.115$$\pm$$0.043 $ & $ 0.17$$\pm$$0.04$$\pm$$0.01 $ & $ 0.18$$\pm$$0.04$$\pm$$0.01$\\        
 $ 0.400$--$0.700 $ & $0.481$ & $58.0 $ & $ 0.520$$\pm$$0.156$$\pm$$0.049 $ & $ 0.06$$\pm$$0.02$$\pm$$0.01 $ & $ 0.07$$\pm$$0.02$$\pm$$0.01$\\        
\hline \hline
\end{tabular}

\caption{ {\em Results on the virtual photon proton asymmetry
\protect$A_1^p$ and the spin
structure function
\protect$g_1^p$ of the proton. The first error is
statistical, the second one is systematic. 
For the evaluation of $g_1^p(Q^2=10$\,GeV$^2)$, 
it has been assumed that $A_1^p$ does not depend on $Q^2$.
\label{table1}}
}
\end{table}

\begin{table}[p] \centering
\begin{tabular}{|l|c|}
\hline \hline
Source of the  error  & $\Delta \Gamma^p_1$\\ 
\hline
Beam polarization  &$0.0057$\\ 
Uncertainty on $F_2$ &$0.0052$\\ 
Extrapolation at low $x$ &$0.0040$\\ 
Target polarization &$0.0039$\\ 
Dilution factor &$0.0034$\\ 
Acceptance variation $\Delta r$ &$0.0030$\\ 
Radiative corrections &$0.0023$\\ 
Neglect of $A_2$ &$0.0017$\\ 
Momentum measurement &$0.0020$\\ 
Uncertainty on $R$ &$0.0018$\\ 
Kinematic resolution&$0.0010$\\ 
Extrapolation at high $x$&$0.0007$\\ 
\hline
Total systematic error  &$0.0113$\\ 
\hline
Statistics &$0.0114$\\ 
\hline \hline
\end{tabular}
\caption {\label{syserr}
{\em Contributions to the error on $\Gamma_1^p$}}
\end{table}


 \begin{figure}[H]
 \begin{center}
 \mbox{ \epsfig{file=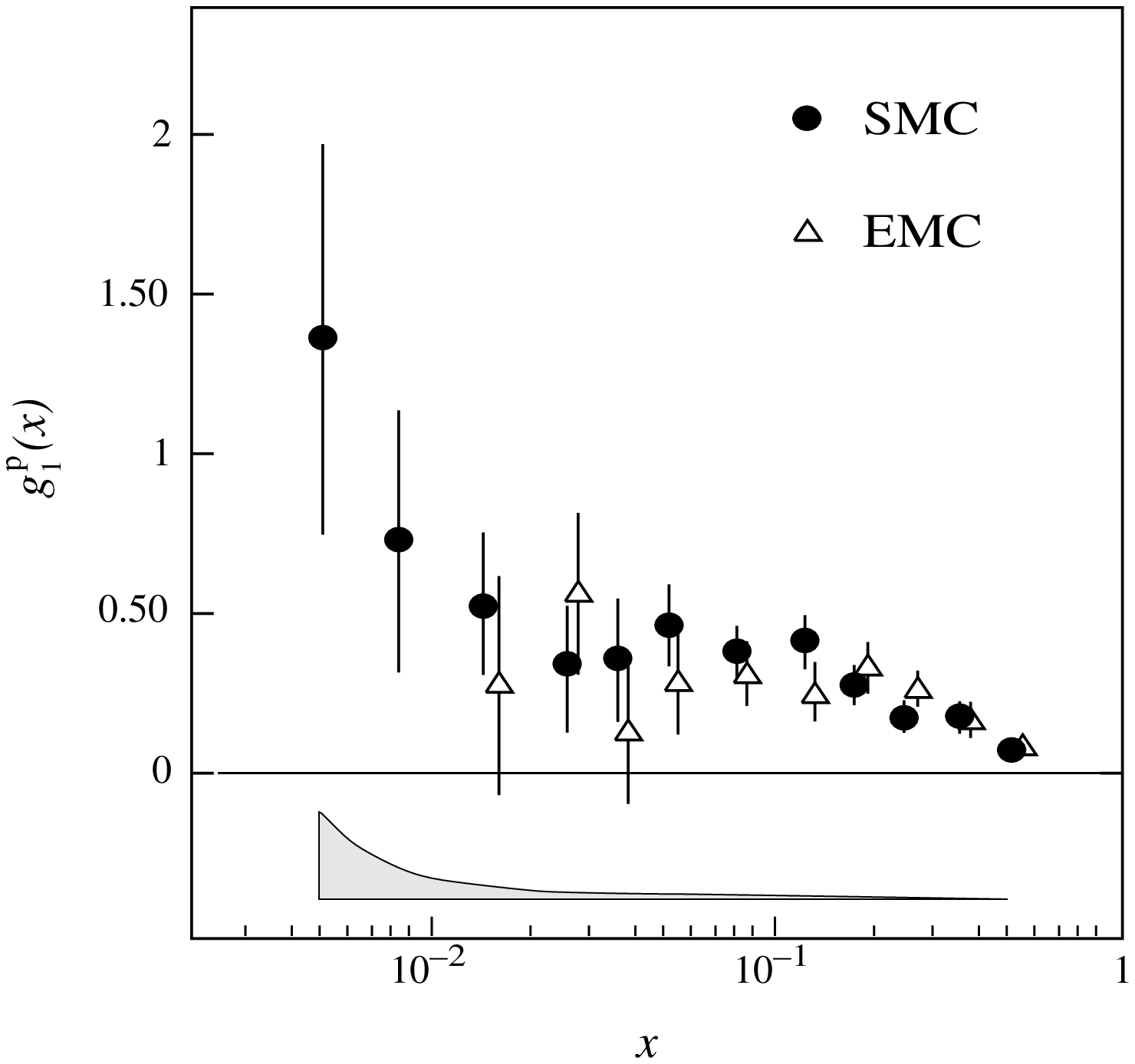,width=10cm}}
 \vspace{5mm}
 \end{center}
 \caption {\em
 The spin dependent structure function $g_1^p(x)$ at the average $Q^2$
 of each $x$ bin (Table~1).  Only statistical errors are shown with the
 data points.  The size of the systematic errors for the SMC data is
 indicated by the shaded area.  The EMC points are reevaluated using
 the NMC $F_2$ parametrization [22]. }
\label{fig3}
 \end{figure}
 \begin{figure}[H]
 \begin{center}
 \mbox{ \epsfig{file=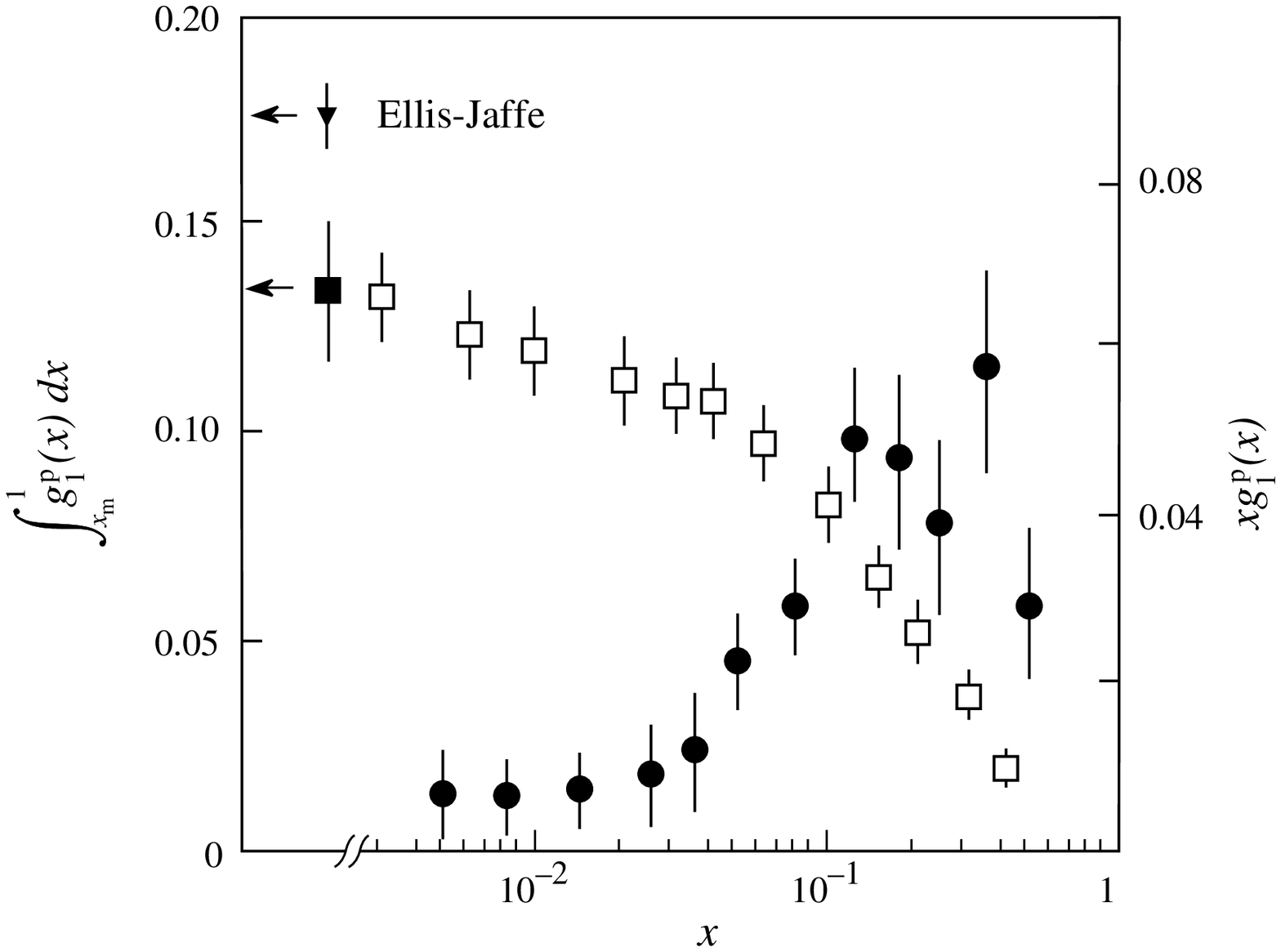,width=12cm}}
 \end{center}
 \caption {\em
 The solid circles (right-hand axis) show the structure function
 $xg_1^p$ as a function of the Bjorken scaling variable $x$, at
 $Q_0^2=10\,GeV^2$. The open boxes (left-hand axis) show
 $\int_{x_{m}}^1g_1^p(x)dx$, where $x_m$ is the value of $x$ at the
 lower edge of each bin. Only statistical errors are shown.  The solid
 square shows our result $\int_0^1g_1^p(x)dx$, with statistical and
 systematic errors combined in quadrature.  Also shown is the
 theoretical prediction by Ellis and Jaffe\,[3].
 }
 \label{fig2}
 \end{figure}

\end{document}